\documentclass{PoS}

\newcommand{\eqn}[1]{(\ref{#1})}
\newcommand{\p}{\partial}

\newcommand{\vev}[1]{\left\langle #1 \right\rangle}

\title{Non-perturbative evaluation of $c_{\rm SW}$ for smeared link clover fermion and Iwasaki gauge action}

\ShortTitle{Non-perturbative evaluation of $c_{\rm SW}$ for smeared link clover fermion}

\author{\speaker{Yusuke Taniguchi}%
\\
Graduate School of Pure and Applied Sciences,
University of Tsukuba,\\
Tsukuba, Ibaraki 305-8571, Japan\\
        E-mail: \email{tanigchi@het.ph.tsukuba.ac.jp}}

\abstract{We performed a rough estimate of the non-perturbative value of
 the clover term coefficient $c_{\rm SW}$ for the APE stout link Wilson
 fermion.
 We varied the number of smearings from $N_{\rm smear}=1$ to $6$ and
 adopted $\beta$ values roughly corresponding to the lattice spacing of
 $0.1$ fm.
 We used the Schr\"odinger functional technique for an evaluation of
 $c_{\rm SW}$ and found that $c_{\rm SW}$ decreases monotonically as we
 increase $N_{\rm smear}$ but has a $10$\% order of deviation from the
 tree level value for $N_{\rm smear}=6$.
}

\FullConference{The 30th International Symposium on Lattice Field Theory\\
		 June 24 - 29,  2012\\
		 Cairns, Australia}

\begin{document}

\section{Introduction}
The $O(a)$ improved Wilson fermion with the smeared link variable 
\cite{Morningstar:2003gk,Capitani:2006ni} is shown to have
several virtues compared with the thin link fermion;
a better scaling behavior, a fewer exceptional configuration
\cite{Durr:2008rw} and
a better chiral behavior \cite{Capitani:2006xd,Durr:2011ap}.

The improvement factor $c_{\rm SW}$ for the clover term is expected to
be smaller than that for the unsmeared link action
\cite{Horsley:2007fw}.
The tree level tadpole improved value seems to be consistent
with the non-perturbative value \cite{Edwards:2008ja}.
Moreover recent studies with highly smeared link action
\cite{Durr:2011ap,Durr:2010vn} adopt the tree level value
$c_{\rm SW}=1$.

In this proceeding we would like to confirm if and how the
non-perturbative value of $c_{\rm SW}$ is close to unity.
We are also interested in the behavior of $c_{\rm SW}$ as we increase
the number of smearings by fixing the lattice spacing.
For this purpose we tried to set $\beta$ to correspond to the lattice
spacing $a=0.1$ fm.

\section{Schr\"odinger functional scheme}
For an evaluation of $c_{\rm SW}$ we make use of the well established
Schr\"odinger functional technique \cite{Luscher:1996ug}.
In this method we measure four kinds of two point functions between the
axial current $A_0$ or the pseudo scalar density $P$ in the bulk and the
``pseudo scalar density'' $O(t=0)$ or $O^{\prime}(t=T)$ at the
temporal boundary
\begin{eqnarray}
&&
f_A(x_0)=-\frac{1}{N_f^2-1}\vev{A_0^a(x_0){O}^a}
,\quad
f_P(x_0)=-\frac{1}{N_f^2-1}\vev{P^a(x_0){O}^a},
\\&&
f_A'(x_0)=+\frac{1}{N_f^2-1}\vev{A_0^a(T-x_0){O'}^a}
,\quad
f_P'(x_0)=-\frac{1}{N_f^2-1}\vev{P^a(T-x_0){O'}^a}.
\end{eqnarray}
The PCAC quark masses are defined in terms of the improved current
\begin{eqnarray}
&&
m(x_0)=r(x_0)+c_As(x_0),\quad
m'(x_0)=r'(x_0)+c_As'(x_0),
\\&&
r(x_0)=\frac{\left(\p_0+\p_0^*\right)f_A(x_0)}{4f_P(x_0)},\quad
s(x_0)=\frac{a\p_0\p_0^*f_P(x_0)}{2f_P(x_0)},
\\&&
r'(x_0)=\frac{\left(\p_0+\p_0^*\right)f_A'(x_0)}{4f_P'(x_0)},\quad
s'(x_0)=\frac{a\p_0\p_0^*f_P'(x_0)}{2f_P'(x_0)}.
\end{eqnarray}
The improved factor $c_A$ for the axial current can be removed by adding
an $O(a^2)$ term
\begin{eqnarray}
M(x_0,y_0)&=&m(x_0)-\frac{m(y_0)-m'(y_0)}{s(y_0)-s'(y_0)}s(x_0)
=r(x_0)-\frac{r(y_0)-r'(y_0)}{s(y_0)-s'(y_0)}s(x_0),
\label{eqn:PCACmass1}
\\
M'(x_0,y_0)&=&m'(x_0)-\frac{m'(y_0)-m(y_0)}{s'(y_0)-s(y_0)}s'(x_0)
=r'(x_0)-\frac{r'(y_0)-r(y_0)}{s'(y_0)-s(y_0)}s'(x_0).
\label{eqn:PCACmass2}
\end{eqnarray}
The massless limit is given by tuning the hopping parameter so that
\begin{eqnarray}
M\left(\frac{T}{2},\frac{T}{4}\right)\to0
\end{eqnarray}
and the non-perturbative $c_{\rm SW}$ is given by the improvement
condition
\begin{eqnarray}
&&
\Delta M
=M\left(\frac{3T}{4},\frac{T}{4}\right)-M'\left(\frac{3T}{4},\frac{T}{4}\right)
\to0
\end{eqnarray}
with the hopping parameter set to its critical value $\kappa_c$.

\section{Simulation setups}
We adopt the Iwasaki gauge action and the improved Wilson fermion action
with the clover term.
The number of flavors is set to $N_f=3$ with degenerate masses, which
shall be tuned to zero.
The APE stout smeared gauge link is used for those in the fermion
action including the clover term.
The smearing parameter is set to
$\rho=0.1$ \cite{Morningstar:2003gk}.
We vary the number of smearings from one to six.
We adopt $8^3\times16$ lattice with the Schr\"odinger functional
boundary condition in the temporal direction \cite{Luscher:1996ug}.

The gauge coupling $\beta$ is tuned so that the lattice spacing becomes
around $0.1$ fm.
Since we would like to know a rough tendency of the improvement
parameter $c_{\rm SW}$ we fix $\beta$ just by guess except for that at
$N_{\rm smear}=6$.
The simulation parameter $\beta$ is given in table
\ref{tab:csw} for each $N_{\rm smear}$.
We checked the lattice spacing is roughly equal to $0.1$ fm by using the
Sommer scale $r_0$ and the $\Omega$ baryon mass input.
The inverse lattice spacing $a^{-1}$ has a tendency to grow up when we
increase $c_{\rm SW}$.

\section{Numerical results}
\begin{figure}
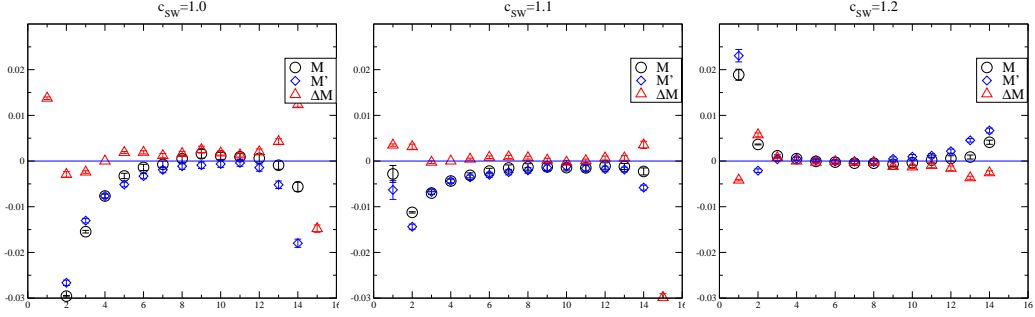

 \begin{center}
  \includegraphics[width=4.5cm]{fig/ns6.rh01.b182.csw100.k12769595.70200.eps}
  \includegraphics[width=4.5cm]{fig/ns6.rh01.b182.csw110.k12629205.37000.eps}
  \includegraphics[width=4.5cm]{fig/ns6.rh01.b182.csw120.k1250889.33000.eps}
  \caption{The PCAC mass $M(x_0,y_0)$, $M'(x_0,y_0)$ and the mass
  difference $\Delta M(x_0,y_0)$ as a function of $x_0$ for
  $N_{\rm smear}=6$, $\beta=1.82$.
  $y_0$ is set to $T/4$.
  Three values of $c_{\rm SW}$ are adopted: $1.0$ (left), $1.1$ (middle)
  and $1.2$ (right).
  The PCAC mass is tuned to be consistent with zero at $x_0=T/2$.
}
  \label{fig:pcacmass}
 \end{center}
\end{figure}

A typical behavior of the PCAC mass $M$ and the mass difference
$\Delta M$ is plotted in figure \ref{fig:pcacmass} for
$N_{\rm smear}=6$ and $\beta=1.82$ at three values of $c_{\rm SW}$.
The PCAC mass difference $\Delta M$ (up triangle) tends to decrease as
we increase $c_{\rm SW}$ and crosses zero around $c_{\rm SW}=1.1$.

\begin{figure}
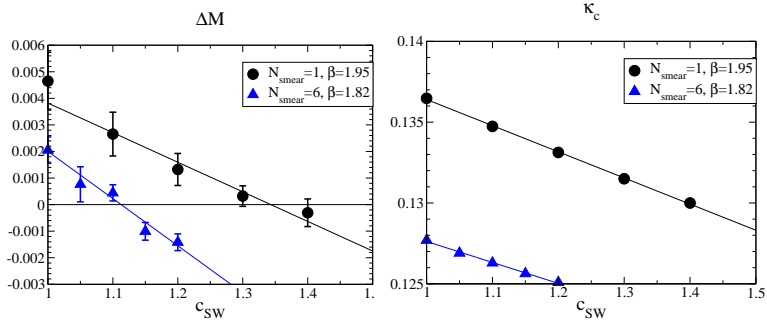

 \begin{center}
  \includegraphics[width=5.0cm]{fig/csw-deltaM.ns1-6.rho01.2.eps}
  \includegraphics[width=5.0cm]{fig/csw-kappa.ns1-6.rho01.eps}
  \caption{The PCAC mass difference $\Delta M(3/4T,T/4)$ (left) and the
  critical hopping parameter $\kappa_c$ (right) as a function
  of $c_{\rm SW}$.
  Two data are plotted with $N_{\rm smear}=1$ (circle) and $6$ (triangle).
}
  \label{fig:csw-deltaM}
 \end{center}
\end{figure}

We plot this behavior in the left panel of figure \ref{fig:csw-deltaM}
for $N_{\rm smear}=1$ and $6$.
Both data can be fitted by a linear function.
The horizontal value where $\Delta M$ crosses zero is the
non-perturbative $c_{\rm SW}$.
The linear fit works well for other $N_{\rm smear}$ and the result
is listed in table \ref{tab:csw}.

Since the hopping parameter is tuned so that the PCAC mass is consistent
with zero it represents the critical $\kappa_c$ at each $c_{\rm SW}$.
$\kappa_c$ can also be fitted linearly as a function of $c_{\rm SW}$ as
is shown in the right panel of figure \ref{fig:csw-deltaM}.
The value of $\kappa_c$ at the non-perturbative $c_{\rm SW}$ is listed
in table \ref{tab:csw} for each $N_{\rm smear}$.
\begin{table}[htb]
\caption{The result for the non-perturbative $c_{\rm SW}$ and the critical
 hopping parameter $\kappa_c$.
 The data for $N_{\rm smear}=0$ is taken from Ref.~\cite{Aoki:2009ix}.
}
\label{tab:csw}
\begin{center}
\begin{tabular}{|c|c|c|c|}
\hline
$N_{\rm smear}$ & $\beta$ & $c_{\rm SW}$ & $\kappa_c$ \\
\hline
$0$ & $1.90$ & $1.715$ & $0.13706$ \\
$1$ & $1.95$ & $1.342(21)$ & $0.13094(34)$ \\
$2$ & $1.93$ & $1.187(18)$ & $0.12871(24)$ \\
$3$ & $1.91$ & $1.155(43)$  & $0.12702(48)$ \\
$4$ & $1.89$ & $1.137(19)$ & $0.12629(23)$ \\
$6$ & $1.87$ & $1.057(20)$ & $0.12634(22)$ \\
$6$ & $1.82$ & $1.1127(96)$ & $0.12612(16)$ \\
\hline
\end{tabular}
\end{center}
\end{table}

\begin{figure}
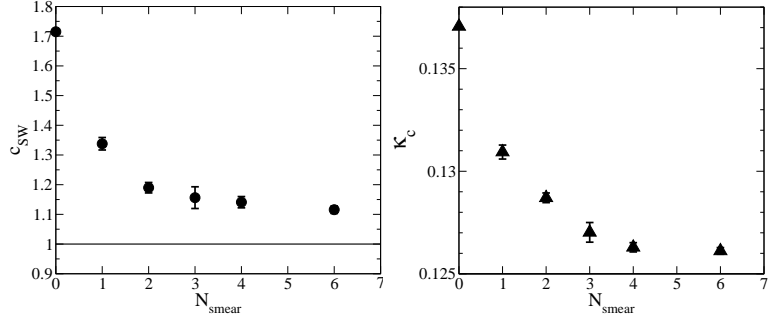

 \begin{center}
  \includegraphics[width=5.0cm]{fig/nsmr-csw.APE.6.eps}
  \includegraphics[width=5.0cm]{fig/nsmr-kappac.APE.4.eps}
  \caption{The non-perturbative $c_{\rm SW}$ (left panel) and the
  critical hopping parameter $\kappa_c$ (right panel) as a function of
  the number of smearings.
}
  \label{fig:csw-nsmr}
 \end{center}
\end{figure}

The non-perturbative $c_{\rm SW}$ is plotted as a function of number of
smearings in the left panel of figure \ref{fig:csw-nsmr}.
$c_{\rm SW}$ decreases monotonically as a function of $N_{\rm smear}$.
We found roughly a $10$\% order of deviation from the tree level value
even at $N_{\rm smear}=6$.
The critical hopping parameter $\kappa_c$ is also given in the right
panel of figure \ref{fig:csw-nsmr}.
The decreasing behavior is almost the same as that of $c_{\rm SW}$ as a
function of the number of smearings.
$\kappa_c$ is very near to the tree level value $1/8$ at
$N_{\rm smear}=6$, which is one of the evidence of the good chiral
behavior of the smeared link action.

\begin{figure}
 \begin{center}
  \includegraphics[width=5.0cm]{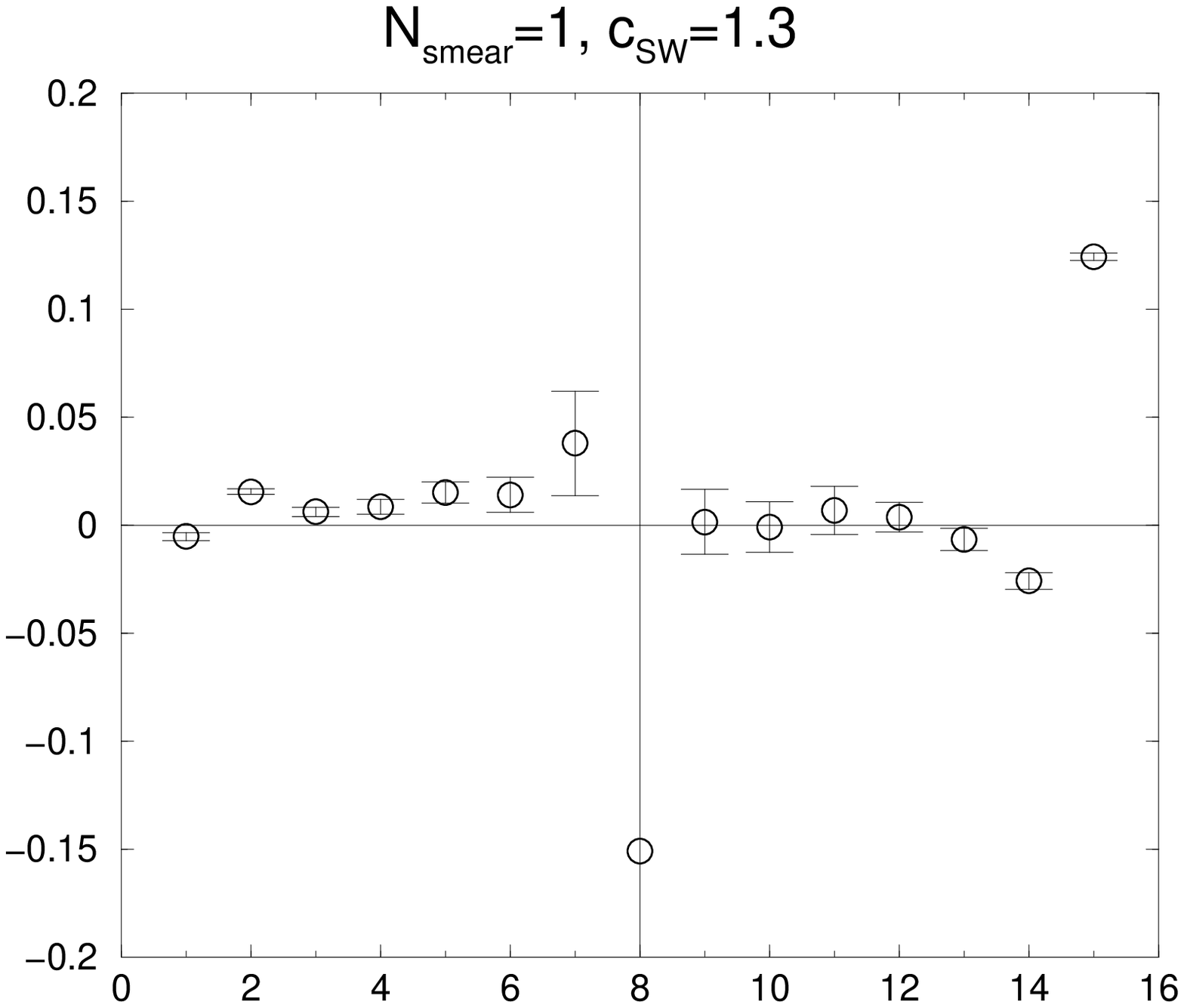}
  \includegraphics[width=5.0cm]{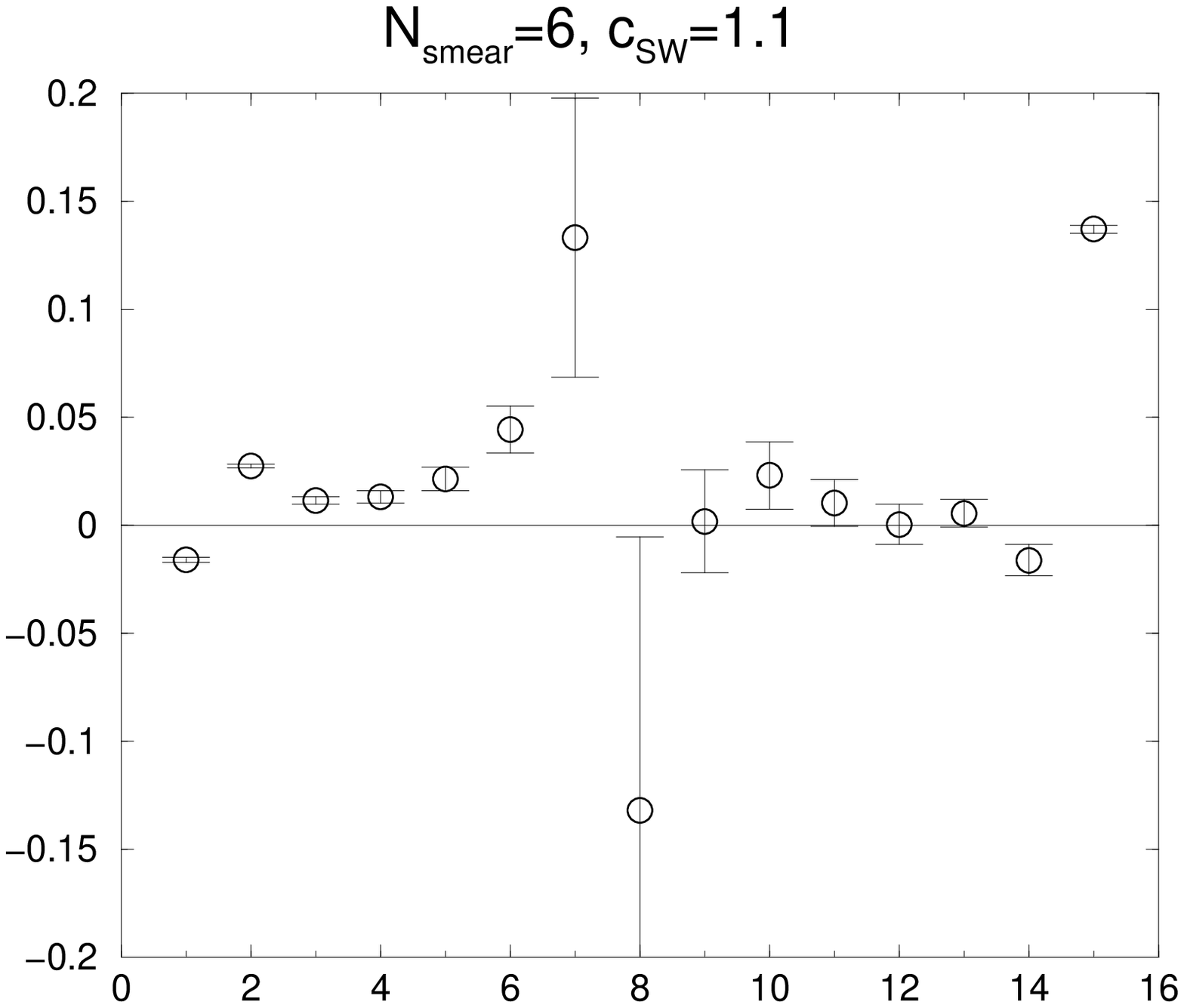}
  \caption{The axial current improvement factor $c_A'$ as a function of
  $x_0$.
  The $c_{\rm SW}$ is set to be nearest to the non-perturbative value.
  The left panel is $N_{\rm smear}=1$, $c_{\rm SW}=1.3$.
  The right panel is $N_{\rm smear}=6$, $c_{\rm SW}=1.1$.
  The hopping parameter is set to $\kappa_c$.
}
  \label{fig:ca-t}
 \end{center}
\end{figure}

\begin{figure}
 \begin{center}
  \includegraphics[width=5.0cm]{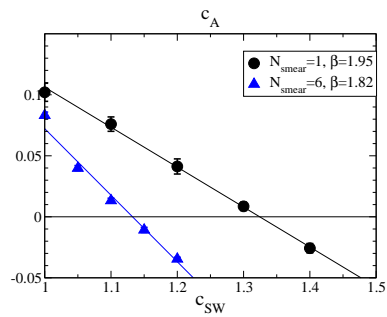}
  \caption{The improvement factor $c_A$ as a function of $c_{\rm SW}$.
  Two data are plotted with $N_{\rm smear}=1$ (circle) and $6$ (triangle).
}
  \label{fig:csw-ca}
 \end{center}
\end{figure}

\begin{figure}
 \begin{center}
  \includegraphics[width=5.0cm]{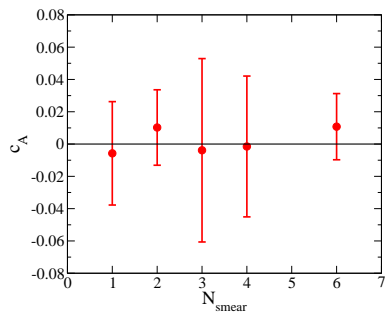}
  \caption{The non-perturbative improvement factor $c_A$ as a function
  of the number of smearings $N_{\rm smear}$.
  All the data are consistent with zero within the statistical error.
}
  \label{fig:nsmr-ca}
 \end{center}
\end{figure}

From \eqn{eqn:PCACmass1} and \eqn{eqn:PCACmass2} a quantity
\begin{eqnarray}
c_A'(x_0)=-\frac{r(x_0)-r'(x_0)}{s(x_0)-s'(x_0)}
\end{eqnarray}
plays a role of the improvement coefficient of the axial vector current.
As can be seen from figure \ref{fig:ca-t} $x_0$ dependence of $c_A'$ is
very flat near the non-perturbative $c_{\rm SW}$ and we are able to
define the improvement factor of the axial current by
\begin{eqnarray}
c_A=c_A'\left(\frac{T}{4}\right).
\end{eqnarray}
This $c_A$ is fitted linearly as a function of $c_{\rm SW}$ as is
shown in figure \ref{fig:csw-ca}.
We evaluate $c_A$ at the non-perturbative $c_{\rm SW}$ and adopt it as
its non-perturbative value.
The results are plotted in figure \ref{fig:nsmr-ca} for each
$N_{\rm smear}$, which turned out to be very small and are consistent
with zero within the statistical error.

\section{Conclusion}
We evaluate the non-perturbative value of the improvement coefficient
$c_{\rm SW}$ of the clover term for the APE smeared link fermion action.
We adopted $N_{\rm smear}=1-6$ as the number of smearings.
The bare coupling $\beta$ is tuned so that the lattice spacing is near
to $a\sim0.1$ fm as possible.
The result is given in table \ref{tab:csw} and figure
\ref{fig:csw-nsmr}.
$c_{\rm SW}$ decreases smoothly as we increase the number of smearings.
However we found a $10$\% deviation from the tree level value even at
$N_{\rm smear}=6$.

As a byproduct we also evaluate the improvement factor $c_A$ of the
axial current, which turned out to be consistent with zero within the
statistical error.

\section*{Acknowledgement}
This work is supported in part by Grants-in-Aid of the Ministry of
Education (Nos. 22540265, 23105701).
This work is in part based on Bridge++ code
(http://suchix.kek.jp/bridge/Lattice-code/).
A part of numerical simulation was performed on T2K-Tsukuba and HA-PACS
\\
(http://www.ccs.tsukuba.ac.jp/CCS/eng/).

\end{document}